\documentstyle[prb,twocolumn,aps, preprint]{revtex}
\begin{document}
\twocolumn[\hsize\textwidth\columnwidth\hsize\csname @twocolumnfalse\endcsname

\title{Magnetic Field Induced Charged Exciton Studies in a
GaAs/Al$_{0.3}$Ga$_{0.7}$As Single Heterojunction}
\author{Yongmin\, Kim$^*$, F. M.\, Munteanu$^{*,\dag}$, C.
H.\, Perry$^{*,\dag}$, D. G.\, Rickel$^*$, J. A.\, Simmons$^{\ddag}$, and J.
L.\, Reno$^{\ddag}$} \address{$^*$National High Magetic Field
Laboratory-Los Alamos National Laboratory, Los Alamos, NM 87545\\
$^{\dag}$Department of Physics, Northeastern University, Boston, MA
02115\\ $^{\ddag}$Sandia National Laboratory, Albuquerque, NM 87185}
\date{\today} \maketitle
\begin{abstract}
The magnetophotoluminescence (MPL) behavior of a
GaAs/Al$_{0.3}$Ga$_{0.7}$As single heterojunction has been
investigated to 60T. We observed negatively charged singlet and
triplet exciton states that are formed at high magnetic fields beyond
the $\nu$=1 quantum Hall state.  The variation of the charged exciton
binding energies are in good agreement with theoretical predictions.
The MPL transition intensities for these states showed intensity
variations (maxima and minima) at the $\nu$=1/3 and 1/5 fractional
quantum Hall (FQH) state as a consequence of a large reduction of
electron-hole screening at these filling factors.
\end{abstract}

]\narrowtext

In the last several years, many magneto-optical investigations have
focused on the $\nu$=1 quantum Hall state due to drastic changes in
its electronic properties in this magnetic field
regime.\cite{1,2,3,4,5,6} In a doped quantum well system, screening
within the two dimensional electron gas (2DEG) prevents the formation
of excitons.  Instead, the 2DEG shows distinct Landau level
transitions in the presence of magnetic fields.  However, whenever the
Fermi energy sweeps from one Landau level to the next, the screening
strength oscillates leading to non-linear behavior in the inter-Landau
transitions observed in photoluminescence (PL).\cite{6} This
non-linear behavior is attributed to a modulation of the hole
self-energy.\cite{7} At $\nu$=1, when all the possible states are
filled, the electron screening is greatly reduced which leads to
formation of the metal-insulator (MI) transition.  Finkelstein {\it
et.  al.}\cite{3} showed that there is a strong correlation between
the MI transition and the appearance of neutral excitons (X$^{0}$)
and negatively charged excitons (X$^{-}$) based on optical
measurements performed on GaAs/AlGaAs quantum wells.  They concluded
that the electrons become less effective in screening at the onset of
the MI transition, allowing the formation of the bound states between
electrons and holes.  More recently, calculations performed by
Whittaker {\it et al.} \cite {8} showed that higher Landau level
corrections are important in obtaining an accurate value for the
binding energy of the charged exciton states.They showed that the
singlet state and not of the triplet state would be the fundamental
state at large magnetic fields. This result contradicts the usual expectation
that the triplet state is the one that becomes the fundamental
state at high fields.  Chapman {\it
et.al.}\cite{9} predicted that quasi two dimensional (2D) systems that
approximate to a biplanar system ({\it e.g.} heterojunctions) are
unlikely to exhibit PL effects due to the charged excitons. This result is
inconsistent with our observations and with those of others.

In this Letter, we report the results of PL measurements on a very
high mobility modulation doped GaAs/Al$_{0.3}$Ga$_{0.7}$As single
heterojunction.  Our polarized MPL measurements enable us to clearly
resolve evidence of both singlet and triplet states of X$^{-}$ that
are formed at high magnetic fields beyond $\nu$=1.  While the
formation of the neutral exciton takes place near the filling factor
$\nu$=1, the charged magneto-excitons are formed at higher fields.
Our measurements also indicate that the singlet state does in
fact remain the fundamental state.  There is no indication in the MPL data of
any cross-over between the triplet and the singlet state at least up
to our high field limit of 58T. The binding energy of the singlet state
(X$_{s}^{-}$) relative to X$^{0}$ is almost independent of magnetic
field, while that of the triplet state (X$_{t}^{-}$) increases slightly
with increasing magnetic field.  Theoretical calculations \cite{7} performed
for the 2D negatively charged excitons predicted an increase of the
binding energy of X$^{-}$ with decreasing well width owing to the
enhancement of the Coulomb interactions induced by confinement.  The
binding energy is expected to increase with magnetic field
due to the shrinkage of the outer electron orbital and an increase in the
Coulomb interaction.  Our experimental results indicate that the
charged exciton binding energies in a single heterojunction more closely
approximate the results expected for a wide quantum well (QW) \cite{7}.  We
also find that the intensity of the PL signal is sensitive to the
formation of the incompressible quantum liquid (IQL) states at the
filling factors $\nu$=1/3 and $\nu$=1/5.

The sample used in this study is a MBE grown
GaAs/Al$_{0.3}$Ga$_{0.7}$As single heterojunction (SHJ) with a dark
electron density of 1.1$\times$10$^{11}cm^{-2}$ and a mobility
greater than 3$\times$10$^{6}cm^{2}$/Vs.  In our PL experiment, the
2DEG density increased to 2.2$\times$10$^{11}cm^{-2}$ under constant
laser illumination.  The high magnetic fields were generated using a
20 T superconducting (SC) magnet and a newly commissioned 60 T
quasi-continuous (QC) magnet, which has a 2-second field duration.  A
$^{4}$He flow cryostat and a $^{3}$He exchange gas system were used to
achieve 2-4 K temperatures in 20T SC magnet and 0.4-4 K in the 60T QC
magnet, respectively.  For PL experiments, a 630nm low power diode
laser was used for the excitation source and a single optical fiber
(600$\mu$m diameter, 0.16 numerical aperture ) technique was employed
to provide both the input excitation light onto sample and the output
PL signal to the spectrometer \cite{10}. The spectroscopic system consisted of
a 300 mm focal length f/4 spectrograph and a charge coupled device
(CCD) detector, which has a fast refresh rate (476Hz) and high quantum
efficiency.  This fast detection system, allowed us to collect
approximately 1000 PL spectra during the 2-second duration of the QC
magnet field pulse.

In Fig.  1, we present two different polarization measurements of PL
spectra taken at the same magnetic field (17 T).  Two side bands are
clearly observed
in the $\sigma$+ (right circularly
polarized, RCP) measurements; these features are not present in the
$\sigma$- (left
circularly polarized, LCP) spectra.  Magnetoresistance data taken
simultaneously with MPL demonstrate that the magnetic field, where
these side peaks first emerge, is just slightly higher than the
$\nu$=1 (B=9.1T) state.  The side features are attributed to charged
exciton transitions due to the charge imbalance between the conduction
and valence bands in the MI transition regime.  In the $\sigma$+
spectra, the peak located 2.1meV below the E0(e1-hh) transition is
first clearly observed at B=13T and is associated with the singlet state
of the charged exciton (X$_{s}^{-}$).  The other peak at 0.6meV below
the E0 transition emerges around 17T. We believe this weaker feature to
be from the PL from the triplet state of the charged exciton
(X$_{t}^{-}$).  The highest energy peak at 1526.3 meV and the weak
peak at 1523 meV that appear in both polarizations are related to the
response of the bulk GaAs free exciton (FX) transition and to an
impurity transition, respectively.

Fig.  2 shows the PL transition energy vs.  magnetic field up to 58T
at T=1.5K. The highest transition (solid line) at low fields is the
0$\rightarrow$0 Landau transition from the 2DEG; this transforms into
the neutral E0 exciton at a field around $\nu$=1.  As previously
reported by Finkelstein {\it et.al.},\cite{4} the PL signal from
charged excitons emerge for filling factors $\nu$$>$1, after the
appearance of the neutral exciton X$^{0}$.  Finkelstein {\it et.al.}
also showed that for large electron density systems, the charged
exciton states are destroyed because of the Coulomb screening of the
free electron gas.  Hence, X$^{-}$ transitions appear only if the
screening of the interaction between a neutral X$^{0}$ and a free
electron is substantially diminished.  With increasing magnetic field,
the screening factor oscillates,\cite{7} reaching a minimum value at
the filling factor $\nu$=1. For a SHJ, photocreated holes are forced
to migrate towards the GaAs flat band region\cite{6} because the
Coulomb repulsion between positive donors and valence holes is larger
than the Coulomb attraction between electrons and holes.  However, at
the MI transition, where the screening effect disappears, holes move
back to the junction region and form a strong bound state due to the
recovered Coulomb attraction between electrons and holes.\cite{6}
Consequently, this dynamical movement of the valence holes in a single
heterojunction enables the formation of the X$^{-}$ in the MI
transition regime.  As this occurs only at $\nu$$<$1 in our
experimental data, we may conclude that the
reduction in the screening factor is not sharp, but rather of an
Anderson type\cite{11} because the electrons are still effective in
screening, although this effect is small.

Our spectral data show two $\sigma$+ charged exciton states.  It
has been pointed out by Palacios {\it et.al.}\cite{12}, the probability of
having
X$^{-}$ in $\sigma$+ polarization is larger than that of having
X$^{-}$ in $\sigma$- polarization, due to the fact that in the
$\nu$$<$1 regime the number of the spin-up electrons in the first
Landau level is much higher than that of the spin-down state.  For
this reason, we expect the number of X$^{-}$ excitons formed by the
capture of a spin up electron by a X$^{0}$ to be much higher than the
number of those formed with a spin down electron.  Shields {\it
et.al.}\cite{13} show that the singlet and
triplet state spin wave functions that can be seen in $\sigma$+
polarization are of the form:
\begin{center}$S_{0}=1/\sqrt{2}(e\uparrow e\downarrow-e\uparrow
e\downarrow)h\uparrow$ and,\\
$T_{0}=1/\sqrt{2}(e\uparrow e\downarrow+e\uparrow
e\downarrow)h\uparrow$~~~~~~~\end{center}
with total spins of (+3/2) for both of them.  These spin wavefunctions
have the lowest possible energies and their formation involves the
presence of a ground state heavy hole state (+3/2).  The other two possible
triplet states are, \begin{center}$T_{-1}=e\downarrow e\downarrow h\uparrow$
 and,\\ $T_{+1}=e\uparrow e\uparrow h\downarrow$~~~~~~~\end{center}
and they will generate a $\sigma$+ and $\sigma$- polarized signals
respectively.
The $T_{-1}$ state can be neglected due to the higher Zeeman energy

The data allow us to determine the magnetic field
dependence of the binding energy of the charged excitons.  The inset
in Fig. 2 shows the binding energies of X$_{s}^{-}$ and X$_{t}^{-}$
transitions relative to that of X$^{0}$.  The binding energy of the
singlet state remains almost constant to 58T (2.1meV), whereas the
binding energy of the triplet state increases from 0.6meV at 17T to
1.2meV at ~58T with a saturation at high magnetic fields.  In general,
this observation may be considered unusual as at very high fields the
triplet state,
in accordance with Hund's rules, has to be the ground state implying
that the singlet and triplet states have to cross each other.
Palacios {\it et.al.}\cite{12} concluded from the result of calculations
performed in the lowest-Landau-level (LLL) approximation,
that the X$^{-}$ ground state will change from the singlet at zero field to
a triplet at high fields, while the singlet state will become unbound in
the limit of strong magnetic fields.
However, more complete calculations performed by Whittaker
and Shields\cite{8} take into consideration both higher Landau
levels and higher energy subbands. Their results lead to a different
conclusion.
For instance, they report that in the case of a 100$\AA$ quantum well (QW), the
cross-over of these two states is not expected to occur until somewhere
around 35
T. As the well width is increased (eg to 300$\AA$), they find that the
two charged exciton transitions show no crossing even at fields as high as 50T.
In our study on a modulation doped SHJ we observe that the difference
in energy between these two negatively charged exciton states stays fixed at
about 1 meV with no
sign of a crossing. This behavior more closely resembles the result found
for
a single-sided doped wide quantum well.\cite{8} As pointed out
above, in the MI transition regime, valence holes move toward the
interface forming a bound state.  However, due to the dynamics of the
system, valence holes can
only remain somewhere near the junction.  For this reason, the
spatially separated electron hole pairs in a SHJ show behavior similar to a
wide quantum
well.  This assumption is in good agreement with the magnitude of the
binding energies
expected for the charged excitons for a wide QW.  The inset in Fig. 2 indicates
that the binding energies of the X$_{t}^{-}$ and X$_{s}^{-}$ at 17T
are of 0.5meV and 2.1meV, respectively. These values are comparable
to those found by Shields et al.\cite{2} of 0.8 meV and 1.9 meV for
X$_{t}^{-}$ and
X$_{s}^{-}$, respectively, at 8T for a wide QW.

Two important aspects emerge from our experimental data.  We find that the
binding energy of the triplet state increases slightly with increasing
magnetic field, whereas the binding energy of the singlet state remains
approximately constant or even decreases.  This behavior can be
understood from the symmetry of the spatial wave
functions for these states.  The singlet spatial wave function is
symmetric, while the
triplet must be antisymmetric if it is to preserve an overall
parity of -1 for the total wave function.  This is equivalent to
saying that in the singlet state, the two electrons are equally
separated from the hole, while in the triplet case, they are located
at different distances from the hole in order to minimize the
repulsion between them.  With increasing magnetic field, the orbit of the outer
electron in the triplet state is more affected by the field
compared with the orbit of the inner electron. The same is true for the
orbits of the two electrons in
the singlet state, which are located to maximize the attraction between
each of them and the hole. Application of a magnetic field shrinks
the orbits of the electron in neutral excitons and the two electrons
in the singlet state but it has a less significant affect on them than
on the shrinkage of the outer electron in the triplet case.  Thus the
reduction in the orbits will lead to an enhancement of the binding
energies that will be different for each of the three particles at hand.

Fig.  3  shows the evolution of the peak intensities for $X^{0}$ and
X$^{-}$ with magnetic field and it can be seen that both the neutral
and charged excitons show non-linear behavior near the filling factors
$\nu$=1/3 and 1/5.  At these filling factors, the $X^{0}$ transition
intensity shows local minima. This behavior was first reported
by Turberfield et al.\cite{14} and was attributed to the
localization of the electrons in these states concomitant with a
reduction of the screening factor.  They found that
 this reduction in the intensity of the neutral exciton is
accompanied by a similar increase in the intensity of the charged
magneto-excitons.  We observe an increase in the triplet state transition
intensity
at the filling factor $\nu$=1/3 and a reduction at $\nu$=1/5. The
intensity of the singlet state transition, on the other hand, increaseas at
$\nu$=1/5 but remains unchanged at
$\nu$=1/3.  In our view, this intensity behavior is due to the reduction of
the free electron orbits at higher magnetic fields. This causes  the
population of the charged X$^{-}$
to increase compared to that of the neutral $X^{0}$, especially in the case
where
the screening effect is small.  The $X_{t}^{-}$
state is more weakly bound compared with $X_{s}^{-}$ state, since the
singlet state remains the lowest energy state.  For this reason, at
$\nu$=1/3, the energy of the $X_{t}^{-}$ state will be lowered more
than the energy of the $X_{s}^{-}$ leading to a an increase in
population of this state; this results in an increase in the
observed PL intensity.  At $\nu$=1/5, the Coulomb interactions for
both neutral and charged excitons will be very strong, such that neither of
these states will experience a significant decrease in energy at this
filling factor.  As a consequence, the population of the singlet state,
which is the fundamental one, will be increased due to
electron localization, leading to the observed peak in the intensity.
It is of note that this anomalous behavior
can still be observed at temperatures as high as T=1.5K.

In conclusion, we have performed MPL spectral measurements on a high
quality low modulation-doped GaAs/Al$_{0.3}$Ga$_{0.7}$As single
heterojunction. The formation of singlet and triplet states of the
X$^{-}$ charged excitons takes place at high magnetic fields beyond the
$\nu$=1 quantum Hall state.
The binding energy of the X$_{s}^{-}$ remains almost constant, whereas
the binding energy of the X$_{t}^{-}$ increases slightly but tends to
saturate with increasing magnetic field.  Our experimental data
support the theoretical prediction of a non-crossover behavior of
these two states in the magnetic fields regime investigated (up to
58T), so that the singlet state remains the fundamental ground state.
The intensity of the $X_{t}^{-}$ transition shows maxima at $\nu$=1/3 and
$\nu$=1/5.
In contrast, the $X_{s}^{-}$ transition shows a minima at $\nu$=1/5 but
little change at $\nu$=1/3. These features can still be observed at
temperatures as high as T=1.5K.

The authors would like to thank A. H. MacDonald for helpful
discussions and gratefully acknowledge the engineers and technicians at
NHMFL-LANL in the operation of the 60T QC magnet.  Work at NHMFL-LANL
is supported by NSF Cooperative Agreement $\#$ DMR-9527035, the
Department of Energy and the State of Florida.  Work at Sandia
National Laboratory is supported by the Department of Energy.

\begin{figure}
\caption{Spin polarized MPL spectra at B=17T and T=1.5K. The RCP
spectrum (solid line) shows $X^{-}$ peaks whereas LCP spectrum does
not show these features.  The highest energy peak is assigned to the
GaAs free exciton and a small peak located between singlet and triplet
states, which appears in both polarization, is an impurity transition
(indicated as I).  Near $\nu$=1/5 state (inset), the singlet state
intensity is comparable to that of the triplet state.}
\label{Fig1}
\end{figure}

\begin{figure}
\caption{Transition energy vs magnetic field at T=1.5K.  The singlet state
appears near $\nu$=1 (9.1T) whereas the triplet state does not appear
until B~16T. The inset shows the binding energies of charged excitons
relative to the neutral exciton.  The binding energy of the singlet
state remains almost constant up to 58T. The triplet state binding
energy increases with increasing magnetic field.}
\label{Fig2}
\end {figure}

\begin{figure}
\caption{MPL transition intensity vs magnetic field at T=1.5K. The MPL
intensities of $X^{0}$ and $X_{t}^{-}$ have peaks near $\nu$=1/3
state, whereas the intensity of the $X_{s}^{-}$ transition remains
constant over that field region.  However, at $\nu$=1/5, the opposite
behavior is observed.  The $X_{s}^{-}$ transition shows a peak in the
intensity while the $X^{0}$ and $X_{t}^{-}$ transition show dips.}
\label{Fig3}
\end{figure}

\end{document}